\documentclass[prl, aps, amssymb, twocolumn, superscriptaddress, notitlepage, longbibliography]{revtex4-2}

\usepackage{booktabs,multirow,tabularx}

\usepackage[margin=1in]{geometry}
\usepackage{amsmath}
\usepackage{amssymb}
\usepackage{amsthm}
\usepackage{amsfonts}
\usepackage{listings}
\usepackage{enumerate}
\usepackage{latexsym}
\usepackage{psfrag}
\usepackage{bm}
\usepackage{graphicx}
\usepackage[caption=false]{subfig}
\usepackage{blkarray}
\usepackage{array}
\usepackage{color}
\usepackage[normalem]{ulem}
\usepackage[colorlinks,linkcolor=blue,citecolor=blue,urlcolor=blue]{hyperref}
\usepackage{dsfont}
\usepackage{ulem}
\usepackage{bbold}
\usepackage{siunitx}
\usepackage{physics}
\usepackage{comment}
\usepackage{mathtools}
\usepackage{sidecap}
\usepackage{stmaryrd}
\usepackage{inputenc }

\newcommand{\beq}{\begin{equation}}
	\newcommand{\eeq}{\end{equation}}
\newcommand{\beqarray}{\begin{eqnarray}}
	\newcommand{\eeqarray}{\end{eqnarray}}

\newcommand{\eq}[1]{Eq.~(\ref{#1})}
\newcommand{\fig}[1]{Fig.~\ref{#1}}

\usepackage{titlesec}
\newcommand{\beginsupplement}{%
        \setcounter{secnumdepth}{3}
        \setcounter{tocdepth}{3} 
        \setcounter{equation}{0}
        \renewcommand{\theequation}{M\arabic{equation}}%
        \setcounter{figure}{0}
        \renewcommand{\thefigure}{M\arabic{figure}}%
        \renewcommand{\thesubsection}{M\arabic{subsection}}
     }

\begin{document}

\title{Non-Hermitian topological ohmmeter}

\author{Viktor K\"{o}nye}
\thanks{These two authors contributed equally}
\affiliation{Leibniz Institute for Solid State and Materials Research,
IFW Dresden, Helmholtzstrasse 20, 01069 Dresden, Germany}
\affiliation{W\"{u}rzburg-Dresden Cluster of Excellence ct.qmat, 01062 Dresden, Germany}

\author{Kyrylo Ochkan}
\thanks{These two authors contributed equally}
\affiliation{Leibniz Institute for Solid State and Materials Research,
IFW Dresden, Helmholtzstrasse 20, 01069 Dresden, Germany}
\affiliation{W\"{u}rzburg-Dresden Cluster of Excellence ct.qmat, 01062 Dresden, Germany}

\author{Anastasiia Chyzhykova}
\affiliation{Leibniz Institute for Solid State and Materials Research,
IFW Dresden, Helmholtzstrasse 20, 01069 Dresden, Germany}
\affiliation{Taras Shevchenko National University of Kyiv, Volodymyrska Street 60, 01033 Kyiv, Ukraine}

\author{Jan Carl Budich}
\affiliation{Department of Physics, TU Dresden, 01062 Dresden, Germany}
\affiliation{W\"{u}rzburg-Dresden Cluster of Excellence ct.qmat, 01062 Dresden, Germany}

\author{Jeroen van den Brink}
\affiliation{Leibniz Institute for Solid State and Materials Research,
IFW Dresden, Helmholtzstrasse 20, 01069 Dresden, Germany}
\affiliation{W\"{u}rzburg-Dresden Cluster of Excellence ct.qmat, 01062 Dresden, Germany}
\affiliation{Department of Physics, TU Dresden, 01062 Dresden, Germany}

\author{Ion Cosma Fulga}
\email{i.c.fulga@ifw-dresden.de}
\affiliation{Leibniz Institute for Solid State and Materials Research,
IFW Dresden, Helmholtzstrasse 20, 01069 Dresden, Germany}
\affiliation{W\"{u}rzburg-Dresden Cluster of Excellence ct.qmat, 01062 Dresden, Germany}

\author{Joseph Dufouleur}
\email{j.dufouleur@ifw-dresden.de}
\affiliation{Leibniz Institute for Solid State and Materials Research,
IFW Dresden, Helmholtzstrasse 20, 01069 Dresden, Germany}
\affiliation{W\"{u}rzburg-Dresden Cluster of Excellence ct.qmat, 01062 Dresden, Germany}
	
\date{\today}

\begin{abstract}

Measuring large electrical resistances forms an essential part of common applications such as insulation testing, but suffers from a fundamental problem:
the larger the resistance, the less sensitive a canonical ohmmeter is.
Here we develop a conceptually different electronic sensor by exploiting the topological properties of non-Hermitian matrices, whose eigenvalues can show an exponential sensitivity to perturbations. 
The ohmmeter is realized in an multi-terminal, linear electric circuit with a non-Hermitian conductance matrix, where the target resistance plays the role of the perturbation.
We inject multiple currents and measure a single voltage in order to directly obtain the value of the resistance.
The relative accuracy of the device increases exponentially with the number of terminals, and
for large resistances outperforms 
a standard measurement by over an order of magnitude. 
Our work paves the way towards leveraging non-Hermitian conductance matrices in high-precision sensing.
\end{abstract}

\maketitle

Small changes typically produce small effects.
This common physical intuition has its roots in mathematics, where according to Weyl's inequality, the spectrum of a Hermitian matrix cannot change by an amount larger than the perturbation.
Non-Hermitian matrices, however, are not constrained in this fashion. 
Instead, a small change can produce a large shift of the spectrum, in some cases even growing exponentially as a function of the matrix dimension.
This counter-intuitive property has recently been proposed as a way of constructing new sensor architectures \cite{Wiersig2014, Hodaei2017, Chen2017, Budich2020}: 
In certain condensed-matter and optical systems, gain and loss may lead to an effectively non-Hermitian description of wavefunction dynamics exhibiting enhanced sensitivity to small parameter changes. 
Specifically, an exponentially enhanced spectral sensitivity with respect to boundary conditions has been predicted \cite{Budich2020} to occur as a consequence of nontrivial topology: 
it is protected by an integer-quantized winding number of the complex spectrum \cite{Gong2018, Borgnia2020, Okuma2020, Bergholtz2021}. 

\begin{figure}[h!]
        \includegraphics[width=8.6cm]{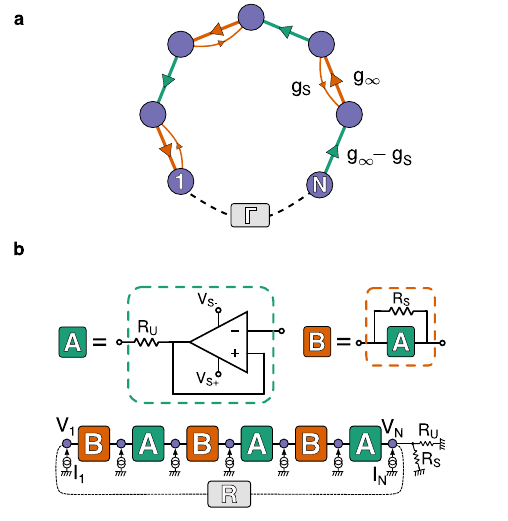}
\caption{
\label{fig:schem} \textbf{Device schematics.} 
 \textbf{(a)} Structure of the conductance matrix \eq{eq:G} for 7 terminals (purple circles), where $g_S$ and $g_\infty$ denote nonreciprocal couplings between adjacent terminals and $\Gamma$ is the reciprocal coupling between the first and the last terminal.
\textbf{(b)} Electric circuit realization using operational amplifiers and resistances where the coupling between the two end terminals is realized by the resistance $R=-1/\Gamma$.
}
\end{figure}

In parallel, it was realized that signatures of nontrivial topology are not unique to condensed-matter systems, but can occur in a variety of other platforms, dubbed metamaterials \cite{Lu2014,Albert2015,Yang2015,Ssstrunk2015,Hu2015,Goldman2016}. 
Their dynamics mimics that of quantum wavefunctions evolving according to the Schr\"{o}dinger equation, allowing for the experimental demonstration of different non-Hermitian topological phenomena \cite{Brandenbourger2019, Lee2019b, Ghatak2020, Weidemann2020, Helbig2020, Xiao2020, Zhang2021a, Zhang2021b, Wang2021, Liu2021, Liang2022}.

Here, starting from these insights and with a focus on applications, 
we build a classical electronic circuit with a non-Hermitian topological conductance matrix \cite{Franca2021, Ochkan2023} that functions as an ohmmeter. 
We consider a multiterminal system connected to a set of current sources and voltmeters that, in linear order, is described by a conductance matrix $\hat{G}$ relating the current vector $\mathbf{I}= (I_1, \dots , I_N$) to the voltage vector $\mathbf{V}= (V_1, \dots ,V_N $). 
We used resistors and operational amplifiers connected in voltage followers configuration to build a circuit associated to a non-Hermitian conductance matrix as shown in \fig{fig:schem}. 
The sensor circuit realizes an electronic counterpart of a non-Hermitian topological system investigated theoretically in the context of tight-binding models in Ref.~\cite{Budich2020} (details in Methods Sec.~\ref{sec:pert}). 
Here, the role of the Hermitian coupling term between the first and the last sites of the tight-binding Hamiltonian is played by the resistor $R$ in \fig{fig:schem}b. 
In this work, we make use of the topological properties of the non-Hermitian $\hat{G}$-matrix to measure the very large resistance $R$ of the device under test (DUT) with enhanced precision.

A standard method to measure large resistances is the constant voltage method where a known (large) voltage is sourced and the current flowing through the DUT is measured. 
The resistance is given by the ratio between the voltage applied on the DUT and the current, the latter being usually determined by measuring the voltage drop on a well-calibrated test resistance.
Such a current vanishes as the resistance $R$ of the DUT increases and high precision measurements of infinitesimal currents are required to achieve a decent measurement of resistances above the $\si{\mega\ohm}$ regime. 
The precision of the current measurement sets the precision of the resistance measurement, which decreases continuously when $R$ increases. 
Unlike the standard method, we consider here a multiple-source device whose sensitivity increases exponentially with the number of terminals.

\begin{figure}[tb]
        \includegraphics[width=8.6cm]{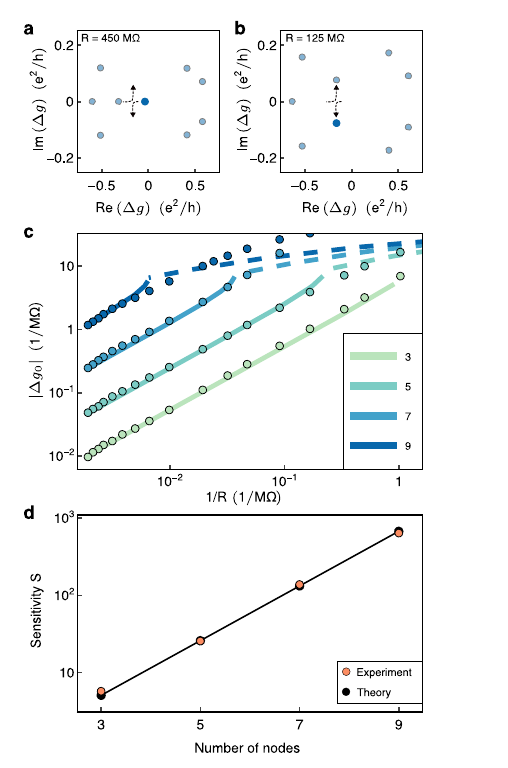}
\caption{\label{Fig. 2}
\textbf{Eigenvalue shift and sensitivity of the device.} 
\textbf{(a)} Eigenspectra of an ideal 9-terminal conductance matrix for the case of separated and 
\textbf{(b)} unseparated eigenvalues. 
For each eigenvalue $g_n$ of ${\hat G}$, $\Delta g = g_n - g_\infty$.
The highlighted eigenvalue corresponds to $\Delta g_0$. 
The dashed arrows indicate how the eigenvalues evolve as $R$ is decreased.
\textbf{(c)} Predicted shift in the conductance eigenvalue $g_0$ as a function of $1/R$ (lines) compared with the experimental data (points) for different numbers of terminals. 
Solid and dashed lines correspond to the region of separated and unseparated eigenvalues respectively.
\textbf{(d)} Predicted sensitivity \eq{eq:S} for different numbers of terminals compared to the experimental values calculated with \eq{eq:mesS} using data presented in panel (c). 
Only the points that correspond to the perturbative regime (separated eigenvalues) were used. 
The error bar is smaller than the symbol size.
}
\end{figure}

The conductance matrix of the electronic circuit in \fig{fig:schem} (for five terminals) is of the form
\begin{equation}
\label{eq:G}
\hat{G}=
  \begin{pmatrix}
    g_\infty - \Gamma & -g_\infty& 0 & 0 & \Gamma\\
    -g_S &  g_\infty & -g_\infty +g_S &0 & 0\\
    0 & 0 &  g_\infty & -g_\infty &0 \\
    0 & 0 & -g_S &  g_\infty & -g_\infty +g_S \\
    \Gamma & 0 & 0 & 0 & g_\infty - \Gamma \\ 
  \end{pmatrix},
\end{equation}
where $g_\infty=1/R_U + 1/R_S$, $\Gamma=-1/R$ and $g_{S}=1/R_{S}$.
$R_U$ is the unidirectional resistance associated to the voltage follower, $R_S$ is a resistance that appears between adjacent terminals in a staggered way, and $R$ is the resistance of the DUT, that connects the first terminal to the last one in our electronic circuit, controlling the boundary condition.
Since we are interested in measuring small variations in the large $R$, we consider the regime where $R \gg R_S,R_U$ so that the resistance is only perturbatively connecting the first terminal to the last one. 
The difference between Eq.~\eqref{eq:G} and the tight-binding Hamiltonian matrix considered in Ref.~\cite{Budich2020} is the $-\Gamma$ at the two ends of the main diagonal (for more information on the non-Hermitian tight-binding model, see Methods Sec.~\ref{sec:pert}). 

We consider the $\hat{G}\mathbf{V_r} = g(R) \mathbf{V_r}$ right eigenvalue problem, where $\mathbf{V_r}$ are the right voltage eigenvectors of the conductance matrix and $g$ is the corresponding eigenvalue (for non-Hermitian matrices left and right eigenvectors are not equivalent, for mathematical details see, e.g. Ref.~\cite{Ashida2020}).
Taking the $R\to\infty$ ($\Gamma=0$) limit, we get the open boundary condition of the non-Hermitian matrix.
In this case, for any odd number of terminals, the matrix is guaranteed to have an eigenvalue equal to $g_0(\infty)=1/R_S+1/R_U\equiv g_\infty$.
In the language of condensed-matter physics this property is a consequence of sublattice symmetry (see Methods Sec.~\ref{sec:pert}).
We focus on the $g_\infty$ eigenvalue below.

For finite resistance $R$, the shift of $g_0(R)$ is given by $\Delta g_0(R) = g_0(R)-g_\infty$ (see Fig.~\ref{Fig. 2}a, b and c).
For large enough $R$, the modulus of $g_0(R)$ is well separated from the modulus of any other eigenvalue (see Fig.~\ref{Fig. 2}a).
We call this the perturbative, or `separated eigenvalues' regime.
Going to smaller $R$ this no longer holds: 
as shown in Fig.~\ref{Fig. 2}a, b, $g_0(R)$ merges with another eigenvalue as $R$ is decreased. 
The pair of eigenvalues then acquire nonzero imaginary parts and have equal absolute values. 
More details on the eigenvalues as a function of $R$ are shown in the Methods Sec.~\ref{sec:eigs}.

At large enough resistances, where the $g_0(R)$ eigenvalue is well separated, we can use perturbation theory \cite{Budich2020} to track the evolution of $\Delta g_0$ (for details see Methods Sec.~\ref{sec:pert}):
\begin{equation}
\label{eq:g0}
     \Delta g_0(R) \approx \frac{\mathbf{V_{l\infty}}^\dag\hat{G}\mathbf{V_{r\infty}}}{\mathbf{V_{l\infty}}^\dag \mathbf{V_{r\infty}}}-g_\infty = \left(-\frac{R_S}{R_U}\right)^{\frac{N-1}{2}} \frac{1}{R}.
\end{equation}
Here, $\mathbf{V_{l\infty}}$ and $\mathbf{V_{r\infty}}$ are the left and right eigenvectors of $\hat{G}$ at $R\to\infty$, and $N$ is the number of terminals.
We also calculate the change $\Delta g_0$ as a function of $1/R$ numerically from $\hat{G}$ as shown in Fig.~\ref{Fig. 2}c with the solid and dashed lines, the solid lines showing where the perturbative results are valid. 
The figure shows that for more terminals the resistances where the perturbative results hold get larger. 
We note that, in parallel to our work, such a shift of the eigenvalues was recently measured in optics \cite{Parto2023} and it was used to measure capacitances in electronic circuits  \cite{Yuan2023}.

Using the definition of Ref~\cite{Budich2020} for the sensitivity $S$ as the change of the eigenvalue with respect to the change in boundary condition we get
\begin{equation}
\label{eq:S}
S=\frac{\dd{g_0}}{\dd{(1/R)}}  = \left(-\frac{R_S}{R_U}\right)^{\frac{N-1}{2}}.
\end{equation}
The expected perturbative values for the sensitivity are shown in Fig.~\ref{Fig. 2}d.
In practice, $R_S$ and $R_U$ may take a large range of values, and could be optimized such as to produce a maximal sensitivity to the target resistance $R$.
For a proof of principle we take here $R_S=\SI{130}{\kilo\ohm}$ and $R_U=\SI{25.5}{\kilo\ohm}$ -- a precise optimization is beyond our present scope.

We connect the first and the last terminal with a resistance $R$ and inject the $\mathbf{I_{r\infty}}\propto \mathbf{V_{r\infty}}$ current eigenvector in the device (see Sec.~\ref{sec:pert} and Sec.~\ref{sec:mes}).
Note that this means multiple current sources are used simultaneously. 
Each of the currents is generated by applying a low frequency AC voltage on a $\SI{1}{\mega\ohm}$ resistor with the source of a lock-in amplifier. Lock-in amplifiers are also used to measure the voltages on the terminals (see Sec.~\ref{sec:mes}). 
Measuring the voltage $V_N$ on the last terminal and dividing by the current on the last terminal $I_N$ enables us to experimentally determine the eigenvalue shift, since (see Sec.~\ref{sec:pert} for details)
\begin{equation}
\label{eq:mesg}
    \frac{V_N}{I_N} \approx \frac{1}{g_0(R)} = \frac{1}{\frac{S}{R}+g_\infty}.
\end{equation}
The measured values for $\Delta g_0(R)$ are shown in Fig.~\ref{Fig. 2}c together with the theoretical prediction. 
The value of $g_\infty$ is determined separately for each point by measuring $V_N$ without connecting the resistance $R$ prior to any finite $R$ measurement, in order to get rid of slow drifts of the voltages, 
and each data point corresponds to an averaging over 30~measurements (see Methods Sec.~\ref{sec:mes} for details). 
There is a very good agreement between the experimental data and the exact eigenvalues for large $R$ values, where the perturbative results hold.

Using the measured value for the voltage and the value for the injected current we can also estimate $S$ as:
\begin{align}
\label{eq:mesS}
 S = R\left(\frac{I_N}{V_N}-g_\infty\right).
\end{align}
We calculate the experimental value of $S$ using this formula for resistances in the range of validity of the perturbative theory (based on the solid lines of Fig.~\ref{Fig. 2}c). 
The results are shown on \fig{Fig. 2}d and they are in good agreement with the theoretical predictions.

\begin{figure}[ht]
        \includegraphics[width=8.6cm]{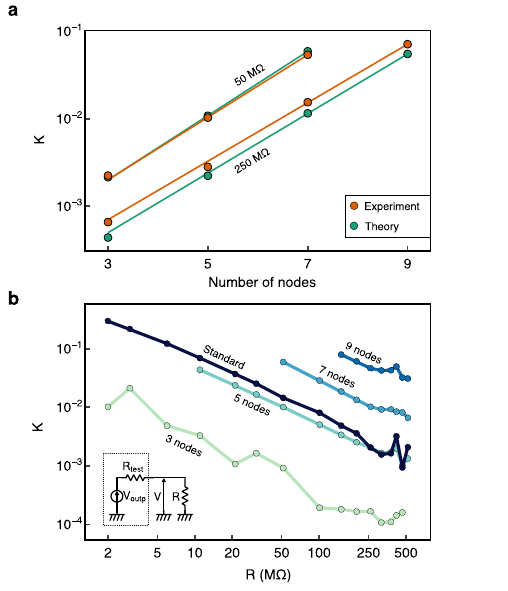}
\caption{\label{Fig. 3}
\textbf{Relative slope K.} 
\textbf{(a)} Relative slope versus different number of nodes for $R \sim \SI{50}{\mega\ohm}$ and for $R \sim \SI{250}{\mega\ohm}$. 
Measurements are realized on a range of resistance corresponding to $10\%$ of $R$  (see Methods Sec.~\ref{sec:mes}). 
Theoretical points correspond to the perturbation theory expectation~\eq{eq:K}. 
Solid lines correspond to linear fits.
\textbf{(b)} Comparison between $K$ values for a standard single-terminal measurement and sensors with different numbers of terminals. The data points for the sensors are obtained in the same way as in \textbf{(a)}. 
The inset shows a standard single-terminal measurement electronic schematic where the dashed line identifies the single-terminal ohmmeter device with 
$R_{\text{test}} = \SI{1}{\mega\ohm}$, $V_{\text{outp}} = \SI{1}{\volt}$.  
}
\end{figure}

For small changes $\Delta R$ of the large target resistance the sensor gives an ohmic relationship between $\Delta R$ and the change in voltage $\Delta V_N$: $\Delta V_N \propto \Delta R$. 
We therefore define the dimensionless relative slope $K$ as 
\begin{equation}
    K = \frac{dV_N}{dR}\frac{R}{V_N}.
\end{equation}
$K$ is independent of the input current $I_1$ and can be expressed as (see Methods Sec.~\ref{sec:pert} for details)
\begin{equation}
\label{eq:K}
    K \approx \frac{I_1}{(S/R+g_\infty)^2R V_N} = \frac{ \dfrac{S}{R}}{\dfrac{S}{R}+g_\infty}.
\end{equation}
$K$ characterizes the relative accuracy of our device.
The relative accuracy of the resistance measurement ($\varepsilon_R=\Delta R/R$) depends on the relative accuracy of the voltage measurement ($\varepsilon_V=\Delta V/V$) as
\begin{equation}
    \varepsilon_R = \frac{\varepsilon_V}{K}.
\end{equation}
Larger $K$ thus improves the accuracy of the device.
From Eq.~\eqref{eq:K} we can see that $K<1$ always holds and from Eq.~\eqref{eq:S} we can see that $K$ grows monotonically as a function of the number of terminals, $N$.
This means that for a given resistance the accuracy can be improved by increasing the number of nodes, provided that the perturbation theory results hold.

The theoretical values for $K$ and the experimentally obtained values are shown on Fig.~\ref{Fig. 3}a as a function of the number of terminals $N$, for which the perturbative theory holds and for two different test resistances ($\SI{50}{\mega\ohm}$ and $\SI{250}{\mega\ohm}$). 
The measurement procedure is described in details in the Methods Sec.~\ref{sec:mes}.

In order to compare the performance of the non-Hermitian ohmmeter with a standard measurement, we measure the resistance $R$ with the circuit shown in the inset of \fig{Fig. 3}b, corresponding to the simplest  configuration of a standard single-terminal circuit. We use the same lock-in amplifiers for sourcing and measuring as for the non-Hermitian ohmmeter. 
In this circuit, the well-calibrated resistance $R_\text{test}$ allows us to determine the current flowing through the resistance $R$ of the DUT, $I=(V_\text{outp}-V)/R_\text{test}$, leading to the trivial voltage divider relation between the measured voltage $V$ and the resistance to be measured $R$: $V=V_{\text{outp}} R/(R+R_\text{test})$.
In the constant voltage mode, $R_\text{test}$ should be as large as possible to maximize the accuracy of the measurement but it should remain much smaller than the resistance $R$ in parallel with the input resistance of the lock-in amplifier ($R_\text{lock}\sim$~10M$\Omega$) to ensure a constant voltage drop on $R$. 
We choose $R_\text{test}=1$~M$\Omega$, a value that fulfilled these two conditions and that corresponds to the polarisation resistance used for the non-Hermitian ohmmeter. 
The results are shown in \fig{Fig. 3}b.
The parameter $K$ shows the same trend as a function of $R$ for all measurements, but can be exponentially increased for a larger number of terminals of the non-Hermitian device.
Thus, we observe that the accuracy of the non-Hermitian ohmmeter outperforms that of the simple, single-terminal measurement starting at $N=7$, and becomes one order of magnitude larger at $N=9$.

\paragraph{\bf Author contributions} 
ICF and JCB conveived the theoretical framework of the project, ICF, JvdB, and JD supervised the project. KO, AC and JD designed the circuit, conducted the measurements and analysed the data. VK performed the analytic calculations and numerical simulations using the idealized and measured conductance matrices to test the feasibility of the device with inputs from JCB, JvdB, and ICF. All authors participated in interpreting the results and writing the manuscript.

\paragraph{\bf Data availability} The data and codes used in this work are available on Zenodo at \cite{zenodocode}.

\paragraph{\bf Acknowledgements} 
This work was supported by the Deutsche Forschungsgemeinschaft (DFG, German Research Foundation) under Germany's Excellence Strategy through the W\"{u}rzburg-Dresden Cluster of Excellence on Complexity and Topology in Quantum Matter -- \emph{ct.qmat} (EXC 2147, project-ids 390858490 and 392019).

\let\oldaddcontentsline\addcontentsline
\renewcommand{\addcontentsline}[3]{}
\bibliography{references.bib}

\begin{thebibliography}{36}%
\makeatletter
\providecommand \@ifxundefined [1]{%
 \@ifx{#1\undefined}
}%
\providecommand \@ifnum [1]{%
 \ifnum #1\expandafter \@firstoftwo
 \else \expandafter \@secondoftwo
 \fi
}%
\providecommand \@ifx [1]{%
 \ifx #1\expandafter \@firstoftwo
 \else \expandafter \@secondoftwo
 \fi
}%
\providecommand \natexlab [1]{#1}%
\providecommand \enquote  [1]{``#1''}%
\providecommand \bibnamefont  [1]{#1}%
\providecommand \bibfnamefont [1]{#1}%
\providecommand \citenamefont [1]{#1}%
\providecommand \href@noop [0]{\@secondoftwo}%
\providecommand \href [0]{\begingroup \@sanitize@url \@href}%
\providecommand \@href[1]{\@@startlink{#1}\@@href}%
\providecommand \@@href[1]{\endgroup#1\@@endlink}%
\providecommand \@sanitize@url [0]{\catcode `\\12\catcode `\$12\catcode
  `\&12\catcode `\#12\catcode `\^12\catcode `\_12\catcode `\%12\relax}%
\providecommand \@@startlink[1]{}%
\providecommand \@@endlink[0]{}%
\providecommand \url  [0]{\begingroup\@sanitize@url \@url }%
\providecommand \@url [1]{\endgroup\@href {#1}{\urlprefix }}%
\providecommand \urlprefix  [0]{URL }%
\providecommand \Eprint [0]{\href }%
\providecommand \doibase [0]{https://doi.org/}%
\providecommand \selectlanguage [0]{\@gobble}%
\providecommand \bibinfo  [0]{\@secondoftwo}%
\providecommand \bibfield  [0]{\@secondoftwo}%
\providecommand \translation [1]{[#1]}%
\providecommand \BibitemOpen [0]{}%
\providecommand \bibitemStop [0]{}%
\providecommand \bibitemNoStop [0]{.\EOS\space}%
\providecommand \EOS [0]{\spacefactor3000\relax}%
\providecommand \BibitemShut  [1]{\csname bibitem#1\endcsname}%
\let\auto@bib@innerbib\@empty
\bibitem [{\citenamefont {Wiersig}(2014)}]{Wiersig2014}%
  \BibitemOpen
  \bibfield  {author} {\bibinfo {author} {\bibfnamefont {J.}~\bibnamefont
  {Wiersig}},\ }\bibfield  {title} {\bibinfo {title} {Enhancing the sensitivity
  of frequency and energy splitting detection by using exceptional points:
  Application to microcavity sensors for single-particle detection},\ }\href
  {https://doi.org/10.1103/PhysRevLett.112.203901} {\bibfield  {journal}
  {\bibinfo  {journal} {Phys. Rev. Lett.}\ }\textbf {\bibinfo {volume} {112}},\
  \bibinfo {pages} {203901} (\bibinfo {year} {2014})}\BibitemShut {NoStop}%
\bibitem [{\citenamefont {Hodaei}\ \emph {et~al.}(2017)\citenamefont {Hodaei},
  \citenamefont {Hassan}, \citenamefont {Wittek}, \citenamefont
  {Garcia-Gracia}, \citenamefont {El-Ganainy}, \citenamefont
  {Christodoulides},\ and\ \citenamefont {Khajavikhan}}]{Hodaei2017}%
  \BibitemOpen
  \bibfield  {author} {\bibinfo {author} {\bibfnamefont {H.}~\bibnamefont
  {Hodaei}}, \bibinfo {author} {\bibfnamefont {A.~U.}\ \bibnamefont {Hassan}},
  \bibinfo {author} {\bibfnamefont {S.}~\bibnamefont {Wittek}}, \bibinfo
  {author} {\bibfnamefont {H.}~\bibnamefont {Garcia-Gracia}}, \bibinfo {author}
  {\bibfnamefont {R.}~\bibnamefont {El-Ganainy}}, \bibinfo {author}
  {\bibfnamefont {D.~N.}\ \bibnamefont {Christodoulides}},\ and\ \bibinfo
  {author} {\bibfnamefont {M.}~\bibnamefont {Khajavikhan}},\ }\bibfield
  {title} {\bibinfo {title} {Enhanced sensitivity at higher-order exceptional
  points},\ }\href {https://doi.org/10.1038/nature23280} {\bibfield  {journal}
  {\bibinfo  {journal} {Nature}\ }\textbf {\bibinfo {volume} {548}},\ \bibinfo
  {pages} {187} (\bibinfo {year} {2017})}\BibitemShut {NoStop}%
\bibitem [{\citenamefont {Chen}\ \emph {et~al.}(2017)\citenamefont {Chen},
  \citenamefont {\"{O}zdemir}, \citenamefont {Zhao}, \citenamefont {Wiersig},\
  and\ \citenamefont {Yang}}]{Chen2017}%
  \BibitemOpen
  \bibfield  {author} {\bibinfo {author} {\bibfnamefont {W.}~\bibnamefont
  {Chen}}, \bibinfo {author} {\bibfnamefont {{\c{S}}.~K.}\ \bibnamefont
  {\"{O}zdemir}}, \bibinfo {author} {\bibfnamefont {G.}~\bibnamefont {Zhao}},
  \bibinfo {author} {\bibfnamefont {J.}~\bibnamefont {Wiersig}},\ and\ \bibinfo
  {author} {\bibfnamefont {L.}~\bibnamefont {Yang}},\ }\bibfield  {title}
  {\bibinfo {title} {Exceptional points enhance sensing in an optical
  microcavity},\ }\href {https://doi.org/10.1038/nature23281} {\bibfield
  {journal} {\bibinfo  {journal} {Nature}\ }\textbf {\bibinfo {volume} {548}},\
  \bibinfo {pages} {192} (\bibinfo {year} {2017})}\BibitemShut {NoStop}%
\bibitem [{\citenamefont {Budich}\ and\ \citenamefont
  {Bergholtz}(2020)}]{Budich2020}%
  \BibitemOpen
  \bibfield  {author} {\bibinfo {author} {\bibfnamefont {J.~C.}\ \bibnamefont
  {Budich}}\ and\ \bibinfo {author} {\bibfnamefont {E.~J.}\ \bibnamefont
  {Bergholtz}},\ }\bibfield  {title} {\bibinfo {title} {Non-hermitian
  topological sensors},\ }\bibfield  {journal} {\bibinfo  {journal} {Physical
  Review Letters}\ }\textbf {\bibinfo {volume} {125}},\ \href
  {https://doi.org/10.1103/physrevlett.125.180403}
  {10.1103/physrevlett.125.180403} (\bibinfo {year} {2020})\BibitemShut
  {NoStop}%
\bibitem [{\citenamefont {Gong}\ \emph {et~al.}(2018)\citenamefont {Gong},
  \citenamefont {Ashida}, \citenamefont {Kawabata}, \citenamefont {Takasan},
  \citenamefont {Higashikawa},\ and\ \citenamefont {Ueda}}]{Gong2018}%
  \BibitemOpen
  \bibfield  {author} {\bibinfo {author} {\bibfnamefont {Z.}~\bibnamefont
  {Gong}}, \bibinfo {author} {\bibfnamefont {Y.}~\bibnamefont {Ashida}},
  \bibinfo {author} {\bibfnamefont {K.}~\bibnamefont {Kawabata}}, \bibinfo
  {author} {\bibfnamefont {K.}~\bibnamefont {Takasan}}, \bibinfo {author}
  {\bibfnamefont {S.}~\bibnamefont {Higashikawa}},\ and\ \bibinfo {author}
  {\bibfnamefont {M.}~\bibnamefont {Ueda}},\ }\bibfield  {title} {\bibinfo
  {title} {Topological phases of non-{H}ermitian systems},\ }\href
  {https://doi.org/10.1103/PhysRevX.8.031079} {\bibfield  {journal} {\bibinfo
  {journal} {Phys. Rev. X}\ }\textbf {\bibinfo {volume} {8}},\ \bibinfo {pages}
  {031079} (\bibinfo {year} {2018})}\BibitemShut {NoStop}%
\bibitem [{\citenamefont {Borgnia}\ \emph {et~al.}(2020)\citenamefont
  {Borgnia}, \citenamefont {Kruchkov},\ and\ \citenamefont
  {Slager}}]{Borgnia2020}%
  \BibitemOpen
  \bibfield  {author} {\bibinfo {author} {\bibfnamefont {D.~S.}\ \bibnamefont
  {Borgnia}}, \bibinfo {author} {\bibfnamefont {A.~J.}\ \bibnamefont
  {Kruchkov}},\ and\ \bibinfo {author} {\bibfnamefont {R.~J.}\ \bibnamefont
  {Slager}},\ }\bibfield  {title} {\bibinfo {title} {{Non-Hermitian Boundary
  Modes and Topology}},\ }\href
  {https://doi.org/10.1103/PHYSREVLETT.124.056802} {\bibfield  {journal}
  {\bibinfo  {journal} {Phys. Rev. Lett.}\ }\textbf {\bibinfo {volume} {124}},\
  \bibinfo {pages} {056802} (\bibinfo {year} {2020})}\BibitemShut {NoStop}%
\bibitem [{\citenamefont {Okuma}\ \emph {et~al.}(2020)\citenamefont {Okuma},
  \citenamefont {Kawabata}, \citenamefont {Shiozaki},\ and\ \citenamefont
  {Sato}}]{Okuma2020}%
  \BibitemOpen
  \bibfield  {author} {\bibinfo {author} {\bibfnamefont {N.}~\bibnamefont
  {Okuma}}, \bibinfo {author} {\bibfnamefont {K.}~\bibnamefont {Kawabata}},
  \bibinfo {author} {\bibfnamefont {K.}~\bibnamefont {Shiozaki}},\ and\
  \bibinfo {author} {\bibfnamefont {M.}~\bibnamefont {Sato}},\ }\bibfield
  {title} {\bibinfo {title} {{Topological Origin of Non-Hermitian Skin
  Effects}},\ }\href {https://doi.org/10.1103/PhysRevLett.124.086801}
  {\bibfield  {journal} {\bibinfo  {journal} {Phys. Rev. Lett.}\ }\textbf
  {\bibinfo {volume} {124}},\ \bibinfo {pages} {086801} (\bibinfo {year}
  {2020})}\BibitemShut {NoStop}%
\bibitem [{\citenamefont {Bergholtz}\ \emph {et~al.}(2021)\citenamefont
  {Bergholtz}, \citenamefont {Budich},\ and\ \citenamefont
  {Kunst}}]{Bergholtz2021}%
  \BibitemOpen
  \bibfield  {author} {\bibinfo {author} {\bibfnamefont {E.~J.}\ \bibnamefont
  {Bergholtz}}, \bibinfo {author} {\bibfnamefont {J.~C.}\ \bibnamefont
  {Budich}},\ and\ \bibinfo {author} {\bibfnamefont {F.~K.}\ \bibnamefont
  {Kunst}},\ }\bibfield  {title} {\bibinfo {title} {Exceptional topology of
  non-hermitian systems},\ }\href
  {https://doi.org/10.1103/RevModPhys.93.015005} {\bibfield  {journal}
  {\bibinfo  {journal} {Rev. Mod. Phys.}\ }\textbf {\bibinfo {volume} {93}},\
  \bibinfo {pages} {015005} (\bibinfo {year} {2021})}\BibitemShut {NoStop}%
\bibitem [{\citenamefont {Lu}\ \emph {et~al.}(2014)\citenamefont {Lu},
  \citenamefont {Joannopoulos},\ and\ \citenamefont
  {Solja{\v{c}}i{\'{c}}}}]{Lu2014}%
  \BibitemOpen
  \bibfield  {author} {\bibinfo {author} {\bibfnamefont {L.}~\bibnamefont
  {Lu}}, \bibinfo {author} {\bibfnamefont {J.~D.}\ \bibnamefont
  {Joannopoulos}},\ and\ \bibinfo {author} {\bibfnamefont {M.}~\bibnamefont
  {Solja{\v{c}}i{\'{c}}}},\ }\bibfield  {title} {\bibinfo {title} {Topological
  photonics},\ }\href {https://doi.org/10.1038/nphoton.2014.248} {\bibfield
  {journal} {\bibinfo  {journal} {Nature Photonics}\ }\textbf {\bibinfo
  {volume} {8}},\ \bibinfo {pages} {821} (\bibinfo {year} {2014})}\BibitemShut
  {NoStop}%
\bibitem [{\citenamefont {Albert}\ \emph {et~al.}(2015)\citenamefont {Albert},
  \citenamefont {Glazman},\ and\ \citenamefont {Jiang}}]{Albert2015}%
  \BibitemOpen
  \bibfield  {author} {\bibinfo {author} {\bibfnamefont {V.~V.}\ \bibnamefont
  {Albert}}, \bibinfo {author} {\bibfnamefont {L.~I.}\ \bibnamefont
  {Glazman}},\ and\ \bibinfo {author} {\bibfnamefont {L.}~\bibnamefont
  {Jiang}},\ }\bibfield  {title} {\bibinfo {title} {Topological properties of
  linear circuit lattices},\ }\bibfield  {journal} {\bibinfo  {journal}
  {Physical Review Letters}\ }\textbf {\bibinfo {volume} {114}},\ \href
  {https://doi.org/10.1103/physrevlett.114.173902}
  {10.1103/physrevlett.114.173902} (\bibinfo {year} {2015})\BibitemShut
  {NoStop}%
\bibitem [{\citenamefont {Yang}\ \emph {et~al.}(2015)\citenamefont {Yang},
  \citenamefont {Gao}, \citenamefont {Shi}, \citenamefont {Lin}, \citenamefont
  {Gao}, \citenamefont {Chong},\ and\ \citenamefont {Zhang}}]{Yang2015}%
  \BibitemOpen
  \bibfield  {author} {\bibinfo {author} {\bibfnamefont {Z.}~\bibnamefont
  {Yang}}, \bibinfo {author} {\bibfnamefont {F.}~\bibnamefont {Gao}}, \bibinfo
  {author} {\bibfnamefont {X.}~\bibnamefont {Shi}}, \bibinfo {author}
  {\bibfnamefont {X.}~\bibnamefont {Lin}}, \bibinfo {author} {\bibfnamefont
  {Z.}~\bibnamefont {Gao}}, \bibinfo {author} {\bibfnamefont {Y.}~\bibnamefont
  {Chong}},\ and\ \bibinfo {author} {\bibfnamefont {B.}~\bibnamefont {Zhang}},\
  }\bibfield  {title} {\bibinfo {title} {Topological acoustics},\ }\bibfield
  {journal} {\bibinfo  {journal} {Physical Review Letters}\ }\textbf {\bibinfo
  {volume} {114}},\ \href {https://doi.org/10.1103/physrevlett.114.114301}
  {10.1103/physrevlett.114.114301} (\bibinfo {year} {2015})\BibitemShut
  {NoStop}%
\bibitem [{\citenamefont {S\"{u}sstrunk}\ and\ \citenamefont
  {Huber}(2015)}]{Ssstrunk2015}%
  \BibitemOpen
  \bibfield  {author} {\bibinfo {author} {\bibfnamefont {R.}~\bibnamefont
  {S\"{u}sstrunk}}\ and\ \bibinfo {author} {\bibfnamefont {S.~D.}\ \bibnamefont
  {Huber}},\ }\bibfield  {title} {\bibinfo {title} {Observation of phononic
  helical edge states in a mechanical topological insulator},\ }\href
  {https://doi.org/10.1126/science.aab0239} {\bibfield  {journal} {\bibinfo
  {journal} {Science}\ }\textbf {\bibinfo {volume} {349}},\ \bibinfo {pages}
  {47} (\bibinfo {year} {2015})}\BibitemShut {NoStop}%
\bibitem [{\citenamefont {Hu}\ \emph {et~al.}(2015)\citenamefont {Hu},
  \citenamefont {Pillay}, \citenamefont {Wu}, \citenamefont {Pasek},
  \citenamefont {Shum},\ and\ \citenamefont {Chong}}]{Hu2015}%
  \BibitemOpen
  \bibfield  {author} {\bibinfo {author} {\bibfnamefont {W.}~\bibnamefont
  {Hu}}, \bibinfo {author} {\bibfnamefont {J.~C.}\ \bibnamefont {Pillay}},
  \bibinfo {author} {\bibfnamefont {K.}~\bibnamefont {Wu}}, \bibinfo {author}
  {\bibfnamefont {M.}~\bibnamefont {Pasek}}, \bibinfo {author} {\bibfnamefont
  {P.~P.}\ \bibnamefont {Shum}},\ and\ \bibinfo {author} {\bibfnamefont
  {Y.}~\bibnamefont {Chong}},\ }\bibfield  {title} {\bibinfo {title}
  {Measurement of a topological edge invariant in a microwave network},\
  }\bibfield  {journal} {\bibinfo  {journal} {Physical Review X}\ }\textbf
  {\bibinfo {volume} {5}},\ \href {https://doi.org/10.1103/physrevx.5.011012}
  {10.1103/physrevx.5.011012} (\bibinfo {year} {2015})\BibitemShut {NoStop}%
\bibitem [{\citenamefont {Goldman}\ \emph {et~al.}(2016)\citenamefont
  {Goldman}, \citenamefont {Budich},\ and\ \citenamefont
  {Zoller}}]{Goldman2016}%
  \BibitemOpen
  \bibfield  {author} {\bibinfo {author} {\bibfnamefont {N.}~\bibnamefont
  {Goldman}}, \bibinfo {author} {\bibfnamefont {J.~C.}\ \bibnamefont
  {Budich}},\ and\ \bibinfo {author} {\bibfnamefont {P.}~\bibnamefont
  {Zoller}},\ }\bibfield  {title} {\bibinfo {title} {Topological quantum matter
  with ultracold gases in optical lattices},\ }\href
  {https://doi.org/10.1038/nphys3803} {\bibfield  {journal} {\bibinfo
  {journal} {Nature Physics}\ }\textbf {\bibinfo {volume} {12}},\ \bibinfo
  {pages} {639} (\bibinfo {year} {2016})}\BibitemShut {NoStop}%
\bibitem [{\citenamefont {Brandenbourger}\ \emph {et~al.}(2019)\citenamefont
  {Brandenbourger}, \citenamefont {Locsin}, \citenamefont {Lerner},\ and\
  \citenamefont {Coulais}}]{Brandenbourger2019}%
  \BibitemOpen
  \bibfield  {author} {\bibinfo {author} {\bibfnamefont {M.}~\bibnamefont
  {Brandenbourger}}, \bibinfo {author} {\bibfnamefont {X.}~\bibnamefont
  {Locsin}}, \bibinfo {author} {\bibfnamefont {E.}~\bibnamefont {Lerner}},\
  and\ \bibinfo {author} {\bibfnamefont {C.}~\bibnamefont {Coulais}},\
  }\bibfield  {title} {\bibinfo {title} {{Non-reciprocal robotic
  metamaterials}},\ }\href {https://doi.org/10.1038/s41467-019-12599-3}
  {\bibfield  {journal} {\bibinfo  {journal} {Nat. Commun.}\ }\textbf {\bibinfo
  {volume} {10}},\ \bibinfo {pages} {1} (\bibinfo {year} {2019})}\BibitemShut
  {NoStop}%
\bibitem [{\citenamefont {Lee}\ and\ \citenamefont {Thomale}(2019)}]{Lee2019b}%
  \BibitemOpen
  \bibfield  {author} {\bibinfo {author} {\bibfnamefont {C.~H.}\ \bibnamefont
  {Lee}}\ and\ \bibinfo {author} {\bibfnamefont {R.}~\bibnamefont {Thomale}},\
  }\bibfield  {title} {\bibinfo {title} {Anatomy of skin modes and topology in
  non-hermitian systems},\ }\bibfield  {journal} {\bibinfo  {journal} {Physical
  Review B}\ }\textbf {\bibinfo {volume} {99}},\ \href
  {https://doi.org/10.1103/physrevb.99.201103} {10.1103/physrevb.99.201103}
  (\bibinfo {year} {2019})\BibitemShut {NoStop}%
\bibitem [{\citenamefont {Ghatak}\ \emph {et~al.}(2020)\citenamefont {Ghatak},
  \citenamefont {Brandenbourger}, \citenamefont {{Van Wezel}},\ and\
  \citenamefont {Coulais}}]{Ghatak2020}%
  \BibitemOpen
  \bibfield  {author} {\bibinfo {author} {\bibfnamefont {A.}~\bibnamefont
  {Ghatak}}, \bibinfo {author} {\bibfnamefont {M.}~\bibnamefont
  {Brandenbourger}}, \bibinfo {author} {\bibfnamefont {J.}~\bibnamefont {{Van
  Wezel}}},\ and\ \bibinfo {author} {\bibfnamefont {C.}~\bibnamefont
  {Coulais}},\ }\bibfield  {title} {\bibinfo {title} {{Observation of
  non-Hermitian topology and its bulk-edge correspondence in an active
  mechanical metamaterial}},\ }\href {https://doi.org/10.1073/PNAS.2010580117}
  {\bibfield  {journal} {\bibinfo  {journal} {PNAS}\ }\textbf {\bibinfo
  {volume} {117}},\ \bibinfo {pages} {29561} (\bibinfo {year}
  {2020})}\BibitemShut {NoStop}%
\bibitem [{\citenamefont {Weidemann}\ \emph {et~al.}(2020)\citenamefont
  {Weidemann}, \citenamefont {Kremer}, \citenamefont {Helbig}, \citenamefont
  {Hofmann}, \citenamefont {Stegmaier}, \citenamefont {Greiter}, \citenamefont
  {Thomale},\ and\ \citenamefont {Szameit}}]{Weidemann2020}%
  \BibitemOpen
  \bibfield  {author} {\bibinfo {author} {\bibfnamefont {S.}~\bibnamefont
  {Weidemann}}, \bibinfo {author} {\bibfnamefont {M.}~\bibnamefont {Kremer}},
  \bibinfo {author} {\bibfnamefont {T.}~\bibnamefont {Helbig}}, \bibinfo
  {author} {\bibfnamefont {T.}~\bibnamefont {Hofmann}}, \bibinfo {author}
  {\bibfnamefont {A.}~\bibnamefont {Stegmaier}}, \bibinfo {author}
  {\bibfnamefont {M.}~\bibnamefont {Greiter}}, \bibinfo {author} {\bibfnamefont
  {R.}~\bibnamefont {Thomale}},\ and\ \bibinfo {author} {\bibfnamefont
  {A.}~\bibnamefont {Szameit}},\ }\bibfield  {title} {\bibinfo {title}
  {{Topological funneling of light}},\ }\href
  {https://doi.org/10.1126/SCIENCE.AAZ8727} {\bibfield  {journal} {\bibinfo
  {journal} {Science}\ }\textbf {\bibinfo {volume} {368}},\ \bibinfo {pages}
  {311} (\bibinfo {year} {2020})}\BibitemShut {NoStop}%
\bibitem [{\citenamefont {Helbig}\ \emph {et~al.}(2020)\citenamefont {Helbig},
  \citenamefont {Hofmann}, \citenamefont {Imhof}, \citenamefont {Abdelghany},
  \citenamefont {Kiessling}, \citenamefont {Molenkamp}, \citenamefont {Lee},
  \citenamefont {Szameit}, \citenamefont {Greiter},\ and\ \citenamefont
  {Thomale}}]{Helbig2020}%
  \BibitemOpen
  \bibfield  {author} {\bibinfo {author} {\bibfnamefont {T.}~\bibnamefont
  {Helbig}}, \bibinfo {author} {\bibfnamefont {T.}~\bibnamefont {Hofmann}},
  \bibinfo {author} {\bibfnamefont {S.}~\bibnamefont {Imhof}}, \bibinfo
  {author} {\bibfnamefont {M.}~\bibnamefont {Abdelghany}}, \bibinfo {author}
  {\bibfnamefont {T.}~\bibnamefont {Kiessling}}, \bibinfo {author}
  {\bibfnamefont {L.~W.}\ \bibnamefont {Molenkamp}}, \bibinfo {author}
  {\bibfnamefont {C.~H.}\ \bibnamefont {Lee}}, \bibinfo {author} {\bibfnamefont
  {A.}~\bibnamefont {Szameit}}, \bibinfo {author} {\bibfnamefont
  {M.}~\bibnamefont {Greiter}},\ and\ \bibinfo {author} {\bibfnamefont
  {R.}~\bibnamefont {Thomale}},\ }\bibfield  {title} {\bibinfo {title}
  {Generalized bulk{\textendash}boundary correspondence in non-{H}ermitian
  topolectrical circuits},\ }\href {https://doi.org/10.1038/s41567-020-0922-9}
  {\bibfield  {journal} {\bibinfo  {journal} {Nat. Phys.}\ }\textbf {\bibinfo
  {volume} {16}},\ \bibinfo {pages} {747} (\bibinfo {year} {2020})}\BibitemShut
  {NoStop}%
\bibitem [{\citenamefont {Xiao}\ \emph {et~al.}(2020)\citenamefont {Xiao},
  \citenamefont {Deng}, \citenamefont {Wang}, \citenamefont {Zhu},
  \citenamefont {Wang}, \citenamefont {Yi},\ and\ \citenamefont
  {Xue}}]{Xiao2020}%
  \BibitemOpen
  \bibfield  {author} {\bibinfo {author} {\bibfnamefont {L.}~\bibnamefont
  {Xiao}}, \bibinfo {author} {\bibfnamefont {T.}~\bibnamefont {Deng}}, \bibinfo
  {author} {\bibfnamefont {K.}~\bibnamefont {Wang}}, \bibinfo {author}
  {\bibfnamefont {G.}~\bibnamefont {Zhu}}, \bibinfo {author} {\bibfnamefont
  {Z.}~\bibnamefont {Wang}}, \bibinfo {author} {\bibfnamefont {W.}~\bibnamefont
  {Yi}},\ and\ \bibinfo {author} {\bibfnamefont {P.}~\bibnamefont {Xue}},\
  }\bibfield  {title} {\bibinfo {title} {{Non-Hermitian bulk–boundary
  correspondence in quantum dynamics}},\ }\href
  {https://doi.org/10.1038/s41567-020-0836-6} {\bibfield  {journal} {\bibinfo
  {journal} {Nat. Phys.}\ }\textbf {\bibinfo {volume} {16}},\ \bibinfo {pages}
  {761} (\bibinfo {year} {2020})}\BibitemShut {NoStop}%
\bibitem [{\citenamefont {Zhang}\ \emph
  {et~al.}(2021{\natexlab{a}})\citenamefont {Zhang}, \citenamefont {Tian},
  \citenamefont {Jiang}, \citenamefont {Lu},\ and\ \citenamefont
  {Chen}}]{Zhang2021a}%
  \BibitemOpen
  \bibfield  {author} {\bibinfo {author} {\bibfnamefont {X.}~\bibnamefont
  {Zhang}}, \bibinfo {author} {\bibfnamefont {Y.}~\bibnamefont {Tian}},
  \bibinfo {author} {\bibfnamefont {J.~H.}\ \bibnamefont {Jiang}}, \bibinfo
  {author} {\bibfnamefont {M.~H.}\ \bibnamefont {Lu}},\ and\ \bibinfo {author}
  {\bibfnamefont {Y.~F.}\ \bibnamefont {Chen}},\ }\bibfield  {title} {\bibinfo
  {title} {{Observation of higher-order non-Hermitian skin effect}},\ }\href
  {https://doi.org/10.1038/s41467-021-25716-y} {\bibfield  {journal} {\bibinfo
  {journal} {Nat. Commun.}\ }\textbf {\bibinfo {volume} {12}},\ \bibinfo
  {pages} {1} (\bibinfo {year} {2021}{\natexlab{a}})}\BibitemShut {NoStop}%
\bibitem [{\citenamefont {Zhang}\ \emph
  {et~al.}(2021{\natexlab{b}})\citenamefont {Zhang}, \citenamefont {Yang},
  \citenamefont {Ge}, \citenamefont {Guan}, \citenamefont {Chen}, \citenamefont
  {Yan}, \citenamefont {Chen}, \citenamefont {Xi}, \citenamefont {Li},
  \citenamefont {Jia}, \citenamefont {Yuan}, \citenamefont {Sun}, \citenamefont
  {Chen},\ and\ \citenamefont {Zhang}}]{Zhang2021b}%
  \BibitemOpen
  \bibfield  {author} {\bibinfo {author} {\bibfnamefont {L.}~\bibnamefont
  {Zhang}}, \bibinfo {author} {\bibfnamefont {Y.}~\bibnamefont {Yang}},
  \bibinfo {author} {\bibfnamefont {Y.}~\bibnamefont {Ge}}, \bibinfo {author}
  {\bibfnamefont {Y.~J.}\ \bibnamefont {Guan}}, \bibinfo {author}
  {\bibfnamefont {Q.}~\bibnamefont {Chen}}, \bibinfo {author} {\bibfnamefont
  {Q.}~\bibnamefont {Yan}}, \bibinfo {author} {\bibfnamefont {F.}~\bibnamefont
  {Chen}}, \bibinfo {author} {\bibfnamefont {R.}~\bibnamefont {Xi}}, \bibinfo
  {author} {\bibfnamefont {Y.}~\bibnamefont {Li}}, \bibinfo {author}
  {\bibfnamefont {D.}~\bibnamefont {Jia}}, \bibinfo {author} {\bibfnamefont
  {S.~Q.}\ \bibnamefont {Yuan}}, \bibinfo {author} {\bibfnamefont {H.~X.}\
  \bibnamefont {Sun}}, \bibinfo {author} {\bibfnamefont {H.}~\bibnamefont
  {Chen}},\ and\ \bibinfo {author} {\bibfnamefont {B.}~\bibnamefont {Zhang}},\
  }\bibfield  {title} {\bibinfo {title} {{Acoustic non-Hermitian skin effect
  from twisted winding topology}},\ }\href
  {https://doi.org/10.1038/s41467-021-26619-8} {\bibfield  {journal} {\bibinfo
  {journal} {Nat. Commun.}\ }\textbf {\bibinfo {volume} {12}},\ \bibinfo
  {pages} {1} (\bibinfo {year} {2021}{\natexlab{b}})}\BibitemShut {NoStop}%
\bibitem [{\citenamefont {Wang}\ \emph {et~al.}(2021)\citenamefont {Wang},
  \citenamefont {Zhang}, \citenamefont {Hua}, \citenamefont {Lei},
  \citenamefont {Lu},\ and\ \citenamefont {Chen}}]{Wang2021}%
  \BibitemOpen
  \bibfield  {author} {\bibinfo {author} {\bibfnamefont {H.}~\bibnamefont
  {Wang}}, \bibinfo {author} {\bibfnamefont {X.}~\bibnamefont {Zhang}},
  \bibinfo {author} {\bibfnamefont {J.}~\bibnamefont {Hua}}, \bibinfo {author}
  {\bibfnamefont {D.}~\bibnamefont {Lei}}, \bibinfo {author} {\bibfnamefont
  {M.}~\bibnamefont {Lu}},\ and\ \bibinfo {author} {\bibfnamefont
  {Y.}~\bibnamefont {Chen}},\ }\bibfield  {title} {\bibinfo {title}
  {{Topological physics of non-Hermitian optics and photonics: a review}},\
  }\href {https://doi.org/10.1088/2040-8986/AC2E15} {\bibfield  {journal}
  {\bibinfo  {journal} {J. Opt.}\ }\textbf {\bibinfo {volume} {23}},\ \bibinfo
  {pages} {123001} (\bibinfo {year} {2021})}\BibitemShut {NoStop}%
\bibitem [{\citenamefont {Liu}\ \emph {et~al.}(2021)\citenamefont {Liu},
  \citenamefont {Shao}, \citenamefont {Ma}, \citenamefont {Zhang},
  \citenamefont {You}, \citenamefont {Wu}, \citenamefont {Xiang}, \citenamefont
  {Cui},\ and\ \citenamefont {Zhang}}]{Liu2021}%
  \BibitemOpen
  \bibfield  {author} {\bibinfo {author} {\bibfnamefont {S.}~\bibnamefont
  {Liu}}, \bibinfo {author} {\bibfnamefont {R.}~\bibnamefont {Shao}}, \bibinfo
  {author} {\bibfnamefont {S.}~\bibnamefont {Ma}}, \bibinfo {author}
  {\bibfnamefont {L.}~\bibnamefont {Zhang}}, \bibinfo {author} {\bibfnamefont
  {O.}~\bibnamefont {You}}, \bibinfo {author} {\bibfnamefont {H.}~\bibnamefont
  {Wu}}, \bibinfo {author} {\bibfnamefont {Y.~J.}\ \bibnamefont {Xiang}},
  \bibinfo {author} {\bibfnamefont {T.~J.}\ \bibnamefont {Cui}},\ and\ \bibinfo
  {author} {\bibfnamefont {S.}~\bibnamefont {Zhang}},\ }\bibfield  {title}
  {\bibinfo {title} {{Non-Hermitian Skin Effect in a Non-Hermitian Electrical
  Circuit}},\ }\href {https://doi.org/10.34133/2021/5608038} {\bibfield
  {journal} {\bibinfo  {journal} {Research}\ }\textbf {\bibinfo {volume}
  {2021}},\ \bibinfo {pages} {1} (\bibinfo {year} {2021})}\BibitemShut
  {NoStop}%
\bibitem [{\citenamefont {Liang}\ \emph {et~al.}(2022)\citenamefont {Liang},
  \citenamefont {Xie}, \citenamefont {Dong}, \citenamefont {Li}, \citenamefont
  {Li}, \citenamefont {Gadway}, \citenamefont {Yi},\ and\ \citenamefont
  {Yan}}]{Liang2022}%
  \BibitemOpen
  \bibfield  {author} {\bibinfo {author} {\bibfnamefont {Q.}~\bibnamefont
  {Liang}}, \bibinfo {author} {\bibfnamefont {D.}~\bibnamefont {Xie}}, \bibinfo
  {author} {\bibfnamefont {Z.}~\bibnamefont {Dong}}, \bibinfo {author}
  {\bibfnamefont {H.}~\bibnamefont {Li}}, \bibinfo {author} {\bibfnamefont
  {H.}~\bibnamefont {Li}}, \bibinfo {author} {\bibfnamefont {B.}~\bibnamefont
  {Gadway}}, \bibinfo {author} {\bibfnamefont {W.}~\bibnamefont {Yi}},\ and\
  \bibinfo {author} {\bibfnamefont {B.}~\bibnamefont {Yan}},\ }\bibfield
  {title} {\bibinfo {title} {Dynamic signatures of non-hermitian skin effect
  and topology in ultracold atoms},\ }\bibfield  {journal} {\bibinfo  {journal}
  {Physical Review Letters}\ }\textbf {\bibinfo {volume} {129}},\ \href
  {https://doi.org/10.1103/physrevlett.129.070401}
  {10.1103/physrevlett.129.070401} (\bibinfo {year} {2022})\BibitemShut
  {NoStop}%
\bibitem [{\citenamefont {Franca}\ \emph {et~al.}(2022)\citenamefont {Franca},
  \citenamefont {K\"onye}, \citenamefont {Hassler}, \citenamefont {van~den
  Brink},\ and\ \citenamefont {Fulga}}]{Franca2021}%
  \BibitemOpen
  \bibfield  {author} {\bibinfo {author} {\bibfnamefont {S.}~\bibnamefont
  {Franca}}, \bibinfo {author} {\bibfnamefont {V.}~\bibnamefont {K\"onye}},
  \bibinfo {author} {\bibfnamefont {F.}~\bibnamefont {Hassler}}, \bibinfo
  {author} {\bibfnamefont {J.}~\bibnamefont {van~den Brink}},\ and\ \bibinfo
  {author} {\bibfnamefont {C.}~\bibnamefont {Fulga}},\ }\bibfield  {title}
  {\bibinfo {title} {Non-{H}ermitian physics without gain or loss: {T}he skin
  effect of reflected waves},\ }\href
  {https://doi.org/10.1103/PhysRevLett.129.086601} {\bibfield  {journal}
  {\bibinfo  {journal} {Phys. Rev. Lett.}\ }\textbf {\bibinfo {volume} {129}},\
  \bibinfo {pages} {086601} (\bibinfo {year} {2022})}\BibitemShut {NoStop}%
\bibitem [{\citenamefont {Ochkan}\ \emph {et~al.}(2023)\citenamefont {Ochkan},
  \citenamefont {Chaturvedi}, \citenamefont {Könye}, \citenamefont {Veyrat},
  \citenamefont {Giraud}, \citenamefont {Mailly}, \citenamefont {Cavanna},
  \citenamefont {Gennser}, \citenamefont {Hankiewicz}, \citenamefont
  {Büchner}, \citenamefont {van~den Brink}, \citenamefont {Dufouleur},\ and\
  \citenamefont {Fulga}}]{Ochkan2023}%
  \BibitemOpen
  \bibfield  {author} {\bibinfo {author} {\bibfnamefont {K.}~\bibnamefont
  {Ochkan}}, \bibinfo {author} {\bibfnamefont {R.}~\bibnamefont {Chaturvedi}},
  \bibinfo {author} {\bibfnamefont {V.}~\bibnamefont {Könye}}, \bibinfo
  {author} {\bibfnamefont {L.}~\bibnamefont {Veyrat}}, \bibinfo {author}
  {\bibfnamefont {R.}~\bibnamefont {Giraud}}, \bibinfo {author} {\bibfnamefont
  {D.}~\bibnamefont {Mailly}}, \bibinfo {author} {\bibfnamefont
  {A.}~\bibnamefont {Cavanna}}, \bibinfo {author} {\bibfnamefont
  {U.}~\bibnamefont {Gennser}}, \bibinfo {author} {\bibfnamefont {E.~M.}\
  \bibnamefont {Hankiewicz}}, \bibinfo {author} {\bibfnamefont
  {B.}~\bibnamefont {Büchner}}, \bibinfo {author} {\bibfnamefont
  {J.}~\bibnamefont {van~den Brink}}, \bibinfo {author} {\bibfnamefont
  {J.}~\bibnamefont {Dufouleur}},\ and\ \bibinfo {author} {\bibfnamefont
  {I.~C.}\ \bibnamefont {Fulga}},\ }\href@noop {} {\bibinfo {title}
  {Observation of non-hermitian topology in a multi-terminal quantum hall
  device}} (\bibinfo {year} {2023}),\ \Eprint
  {https://arxiv.org/abs/2305.18674} {arXiv:2305.18674 [cond-mat.mes-hall]}
  \BibitemShut {NoStop}%
\bibitem [{\citenamefont {Ashida}\ \emph {et~al.}(2020)\citenamefont {Ashida},
  \citenamefont {Gong},\ and\ \citenamefont {Ueda}}]{Ashida2020}%
  \BibitemOpen
  \bibfield  {author} {\bibinfo {author} {\bibfnamefont {Y.}~\bibnamefont
  {Ashida}}, \bibinfo {author} {\bibfnamefont {Z.}~\bibnamefont {Gong}},\ and\
  \bibinfo {author} {\bibfnamefont {M.}~\bibnamefont {Ueda}},\ }\bibfield
  {title} {\bibinfo {title} {Non-{H}ermitian physics},\ }\href
  {https://doi.org/10.1080/00018732.2021.1876991} {\bibfield  {journal}
  {\bibinfo  {journal} {Adv. Phys.}\ }\textbf {\bibinfo {volume} {69}},\
  \bibinfo {pages} {249} (\bibinfo {year} {2020})}\BibitemShut {NoStop}%
\bibitem [{\citenamefont {Parto}\ \emph {et~al.}(2023)\citenamefont {Parto},
  \citenamefont {Leefmans}, \citenamefont {Williams},\ and\ \citenamefont
  {Marandi}}]{Parto2023}%
  \BibitemOpen
  \bibfield  {author} {\bibinfo {author} {\bibfnamefont {M.}~\bibnamefont
  {Parto}}, \bibinfo {author} {\bibfnamefont {C.}~\bibnamefont {Leefmans}},
  \bibinfo {author} {\bibfnamefont {J.}~\bibnamefont {Williams}},\ and\
  \bibinfo {author} {\bibfnamefont {A.}~\bibnamefont {Marandi}},\ }\href@noop
  {} {\bibinfo {title} {Enhanced sensitivity via non-hermitian topology}}
  (\bibinfo {year} {2023}),\ \Eprint {https://arxiv.org/abs/2305.03282}
  {arXiv:2305.03282 [physics.optics]} \BibitemShut {NoStop}%
\bibitem [{\citenamefont {Yuan}\ \emph {et~al.}(2023)\citenamefont {Yuan},
  \citenamefont {Zhang}, \citenamefont {Zhou}, \citenamefont {Wang},
  \citenamefont {Pan}, \citenamefont {Feng}, \citenamefont {Sun},\ and\
  \citenamefont {Zhang}}]{Yuan2023}%
  \BibitemOpen
  \bibfield  {author} {\bibinfo {author} {\bibfnamefont {H.}~\bibnamefont
  {Yuan}}, \bibinfo {author} {\bibfnamefont {W.}~\bibnamefont {Zhang}},
  \bibinfo {author} {\bibfnamefont {Z.}~\bibnamefont {Zhou}}, \bibinfo {author}
  {\bibfnamefont {W.}~\bibnamefont {Wang}}, \bibinfo {author} {\bibfnamefont
  {N.}~\bibnamefont {Pan}}, \bibinfo {author} {\bibfnamefont {Y.}~\bibnamefont
  {Feng}}, \bibinfo {author} {\bibfnamefont {H.}~\bibnamefont {Sun}},\ and\
  \bibinfo {author} {\bibfnamefont {X.}~\bibnamefont {Zhang}},\ }\bibfield
  {title} {\bibinfo {title} {Non-hermitian topolectrical circuit sensor with
  high sensitivity},\ }\bibfield  {journal} {\bibinfo  {journal} {Advanced
  Science}\ }\href {https://doi.org/10.1002/advs.202301128}
  {10.1002/advs.202301128} (\bibinfo {year} {2023})\BibitemShut {NoStop}%
\bibitem [{\citenamefont {K\"{o}nye}\ \emph {et~al.}(2023)\citenamefont
  {K\"{o}nye}, \citenamefont {Ochkan}, \citenamefont {Chyzhykova},
  \citenamefont {Budich}, \citenamefont {van~den Brink}, \citenamefont
  {Fulga},\ and\ \citenamefont {Dufouleur}}]{zenodocode}%
  \BibitemOpen
  \bibfield  {author} {\bibinfo {author} {\bibfnamefont {V.}~\bibnamefont
  {K\"{o}nye}}, \bibinfo {author} {\bibfnamefont {K.}~\bibnamefont {Ochkan}},
  \bibinfo {author} {\bibfnamefont {A.}~\bibnamefont {Chyzhykova}}, \bibinfo
  {author} {\bibfnamefont {J.~C.}\ \bibnamefont {Budich}}, \bibinfo {author}
  {\bibfnamefont {J.}~\bibnamefont {van~den Brink}}, \bibinfo {author}
  {\bibfnamefont {I.~C.}\ \bibnamefont {Fulga}},\ and\ \bibinfo {author}
  {\bibfnamefont {J.}~\bibnamefont {Dufouleur}},\ }\bibfield  {title} {\bibinfo
  {title} {Non-hermitian topological ohmmeter},\ }\bibfield  {journal}
  {\bibinfo  {journal} {Zenodo}\ }\href
  {https://doi.org/10.5281/zenodo.8268147} {10.5281/zenodo.8268147} (\bibinfo
  {year} {2023})\BibitemShut {NoStop}%
\bibitem [{\citenamefont {Su}\ \emph {et~al.}(1979)\citenamefont {Su},
  \citenamefont {Schrieffer},\ and\ \citenamefont {Heeger}}]{Su1979}%
  \BibitemOpen
  \bibfield  {author} {\bibinfo {author} {\bibfnamefont {W.~P.~W.}\
  \bibnamefont {Su}}, \bibinfo {author} {\bibfnamefont {J.~R.}\ \bibnamefont
  {Schrieffer}},\ and\ \bibinfo {author} {\bibfnamefont {A.~J.}\ \bibnamefont
  {Heeger}},\ }\bibfield  {title} {\bibinfo {title} {{Solitons in
  polyacetylene}},\ }\href {https://doi.org/10.1103/PhysRevLett.42.1698}
  {\bibfield  {journal} {\bibinfo  {journal} {Phys. Rev. Lett.}\ }\textbf
  {\bibinfo {volume} {42}},\ \bibinfo {pages} {1698} (\bibinfo {year}
  {1979})}\BibitemShut {NoStop}%
\bibitem [{\citenamefont {Lieu}(2018)}]{Lieu2018}%
  \BibitemOpen
  \bibfield  {author} {\bibinfo {author} {\bibfnamefont {S.}~\bibnamefont
  {Lieu}},\ }\bibfield  {title} {\bibinfo {title} {Topological phases in the
  non-hermitian su-schrieffer-heeger model},\ }\bibfield  {journal} {\bibinfo
  {journal} {Physical Review B}\ }\textbf {\bibinfo {volume} {97}},\ \href
  {https://doi.org/10.1103/physrevb.97.045106} {10.1103/physrevb.97.045106}
  (\bibinfo {year} {2018})\BibitemShut {NoStop}%
\bibitem [{\citenamefont {Hatano}\ and\ \citenamefont
  {Nelson}(1996)}]{Hatano1996}%
  \BibitemOpen
  \bibfield  {author} {\bibinfo {author} {\bibfnamefont {N.}~\bibnamefont
  {Hatano}}\ and\ \bibinfo {author} {\bibfnamefont {D.~R.}\ \bibnamefont
  {Nelson}},\ }\bibfield  {title} {\bibinfo {title} {Localization transitions
  in non-{H}ermitian quantum mechanics},\ }\href
  {https://doi.org/10.1103/physrevlett.77.570} {\bibfield  {journal} {\bibinfo
  {journal} {Phys. Rev. Lett.}\ }\textbf {\bibinfo {volume} {77}},\ \bibinfo
  {pages} {570} (\bibinfo {year} {1996})}\BibitemShut {NoStop}%
\bibitem [{\citenamefont {Kunst}\ \emph {et~al.}(2017)\citenamefont {Kunst},
  \citenamefont {Trescher},\ and\ \citenamefont {Bergholtz}}]{Kunst2017}%
  \BibitemOpen
  \bibfield  {author} {\bibinfo {author} {\bibfnamefont {F.~K.}\ \bibnamefont
  {Kunst}}, \bibinfo {author} {\bibfnamefont {M.}~\bibnamefont {Trescher}},\
  and\ \bibinfo {author} {\bibfnamefont {E.~J.}\ \bibnamefont {Bergholtz}},\
  }\bibfield  {title} {\bibinfo {title} {Anatomy of topological surface states:
  Exact solutions from destructive interference on frustrated lattices},\
  }\href {https://doi.org/10.1103/PhysRevB.96.085443} {\bibfield  {journal}
  {\bibinfo  {journal} {Phys. Rev. B}\ }\textbf {\bibinfo {volume} {96}},\
  \bibinfo {pages} {085443} (\bibinfo {year} {2017})}\BibitemShut {NoStop}%
\bibitem [{\citenamefont {Kunst}\ \emph {et~al.}(2018)\citenamefont {Kunst},
  \citenamefont {Edvardsson}, \citenamefont {Budich},\ and\ \citenamefont
  {Bergholtz}}]{Kunst2018}%
  \BibitemOpen
  \bibfield  {author} {\bibinfo {author} {\bibfnamefont {F.~K.}\ \bibnamefont
  {Kunst}}, \bibinfo {author} {\bibfnamefont {E.}~\bibnamefont {Edvardsson}},
  \bibinfo {author} {\bibfnamefont {J.~C.}\ \bibnamefont {Budich}},\ and\
  \bibinfo {author} {\bibfnamefont {E.~J.}\ \bibnamefont {Bergholtz}},\
  }\bibfield  {title} {\bibinfo {title} {Biorthogonal bulk-boundary
  correspondence in non-hermitian systems},\ }\href
  {https://doi.org/10.1103/PhysRevLett.121.026808} {\bibfield  {journal}
  {\bibinfo  {journal} {Phys. Rev. Lett.}\ }\textbf {\bibinfo {volume} {121}},\
  \bibinfo {pages} {026808} (\bibinfo {year} {2018})}\BibitemShut {NoStop}%
\end{thebibliography}%
\let\addcontentsline\oldaddcontentsline

\clearpage


\beginsupplement
\section*{Methods}

\subsection{non-Hermitian SSH model and perturbation theory results} \label{sec:pert}

The Eq~\eqref{eq:G} conductance matrix is equivalent to the Hamiltonian matrix of a non-Hermitian Su-Schrieffer-Heeger (SSH) model \cite{Su1979,Lieu2018}.
The only difference is the $\Gamma$ present on the diagonal.
The model has non-reciprocal staggered hoppings and $g_\infty$ plays the role of the onsite potential. Hoppings to the right are $-g_S$ and $0$, hoppings to the left are $-g_\infty$ and $g_S-g_\infty$, and $\Gamma$ is the hopping parameter between the first and last site.
Ignoring the diagonal and setting $\Gamma=0$ the $\hat{G}$ matrix has sublattice symmetry $\hat{C}^\dag \hat{G} \hat{C} = -\hat{G}$ with $\hat{C}={\rm diag}(1,-1,1,-1,...)$.
By choosing the length $N$ of the chain to be odd, this symmetry ensures that there is always a zero mode.
Thus, with constant diagonal term and in the $\Gamma=0$ limit, the matrix will always have an eigenvalue $g_0(\infty)=g_\infty$.

Two different limits exist depending on the values of $R_S$ and $R_U$. When $R_S\ll R_U$ the staggering dominates and the system is in a Su–Schrieffer–Heeger model \cite{Su1979} like phase. 
When $R_S\gg R_U$ the unidirectionality dominates and the system is in a Hatano-Nelson model \cite{Hatano1996} like phase. 
In our device, we choose a region in between these two limiting cases, closer to the Hatano-Nelson phase.

Now we show the derivation of the perturbative expressions presented in the main text.
We take the Eq.~\eqref{eq:G} matrix with $R\to\infty$.
The right and left eigenvectors corresponding to the $g_0(\infty)$ eigenvalue can be calculated exactly as~\cite{Kunst2017,Kunst2018}.

\begin{align}
\mathbf{V_{r\infty}}&= V_r\begin{pmatrix} (-R_S/R_U)^{\frac{N-1}{2}} \\ 0 \\ (-R_S/R_U)^{\frac{N-3}{2}} \\ 0 \\ \vdots \\ 0 \\ -R_S/R_U \\ 0 \\ 1 \end{pmatrix} , \\
\mathbf{V_{l\infty}}^\dag&= V_l\begin{pmatrix} 0, & 0, & \dots & 0, & 1 \end{pmatrix} ,
\end{align}
where $V_r$ and $V_l$ are normalizations in the units of voltage.

Using these eigenvectors the perturbative eigenvalue $g_0(R)$ can be calculated as per Eq.~\eqref{eq:g0}.

Injecting the right eigenvectors as currents of the form (we use the normalization where the largest current is $I_1$)
\begin{equation}\label{eq:analytic_current_vec}
    \mathbf{I_{r\infty}} = I_1\begin{pmatrix} 1 \\ 0 \\ -R_U/R_S \\ 0 \\ \vdots \\ 0 \\ (-R_U/R_S)^{\frac{N-3}{2}} \\ 0 \\ (-R_U/R_S)^{\frac{N-1}{2}} \end{pmatrix} \propto \mathbf{V_{r\infty}},
\end{equation}
the voltage on the last terminal can be expressed as \cite{Budich2020}
\begin{equation}
    V_N = \left(\hat{R}\mathbf{I_{r\infty}}\right)_N 
\end{equation}

where $\hat{R}=\hat{G}^{-1}$.
For three terminals this is
\begin{align}
\nonumber
&\hat{R}=\\
&\begin{pmatrix}
    \frac{R_U(RR_U+RR_S+R_UR_S)}{R_U^2+RR_U+RR_S+R_UR_S} & \frac{R_U(RR_U+RR_S+R_UR_S)}{R_U^2+RR_U+RR_S+R_UR_S}&  \frac{R_UR_S}{R_U+R_S}\\[6pt]
    \frac{R_U^2(R+R_S)}{R_U^2+RR_U+RR_S+R_UR_S} &  \frac{R_U(RR_U+RR_S+2R_UR_S)}{R_U^2+RR_U+RR_S+R_UR_S} & \frac{R_UR_S}{R_U+R_S} \\[6pt]
       \frac{R_U^2R_S}{R_U^2+RR_U+RR_S+R_UR_S} & \frac{R_U^2R_S}{R_U^2+RR_U+RR_S+R_UR_S} &  \frac{R_UR_S}{R_U+R_S}
  \end{pmatrix}.
  \end{align}
The reflection matrix has an eigenvalue obeying $r_0(R)=1/g_0(R)$, which can be expressed perturbatively as
\begin{equation}
    r_0(R) \approx \frac{\mathbf{I_{l\infty}}^\dag\hat{R}\mathbf{I_{r\infty}}}{\mathbf{I_{l\infty}}^\dag\mathbf{I_{r\infty}}},
\end{equation}
and since only the last element of $\mathbf{I_{l\infty}}^\dag$ is non-zero, the voltage on the last terminal can be expressed as
\begin{equation}
    V_N = \frac{I_N}{g_0(R)},
\end{equation}
where 
\begin{equation}
I_N = I_1\left(-\frac{R_U}{R_S}\right)^{\frac{N-1}{2}}=I_1/S.
\end{equation}
Using this formula for the voltage we can obtain Eqs.~\eqref{eq:mesg}, \eqref{eq:mesS} and \eqref{eq:K} of the main text.

\subsection{Conductance matrix eigenspectra} \label{sec:eigs}

\begin{figure}[ht]
        \includegraphics[width=8.6cm]{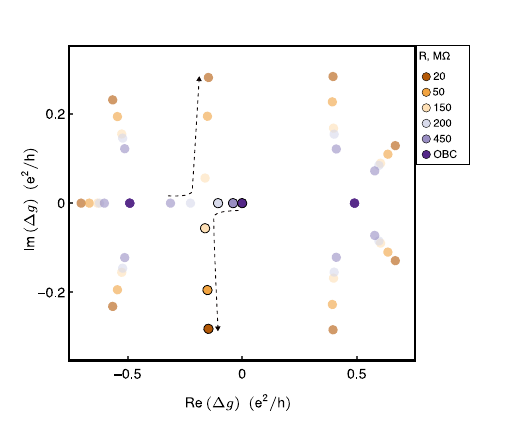}
\caption{\label{fig:SI:spectrum}
\textbf{Conductance matrix eigenspectra} for various values of resistances $R$ between the first and the last node using the idealized 9 terminal conductance matrix in Eq.~\eqref{eq:G}. 
The highlighted eigenvalue is $\Delta g_0$.}
\end{figure}

Figure \fig{fig:SI:spectrum} shows the spectrum of the conductance matrix for additional values of $R$ (compare with \fig{Fig. 2}a and \fig{Fig. 2}b). 
The perturbation theory results hold while the eigenvalue $g_0(R)$ is far away from the next eigenvalue, which we label $g_1$.
For 9 terminals this is true for resistances $R \gg \SI{200}{\mega\ohm}$.

\subsection{Measurement details and procedure} \label{sec:mes}

We use AD 823A FET input operational amplifiers for building the electronic circuit.
The voltages are measured with SR 830 lock-in amplifiers, with an input impedance of about 10~M$\Omega$. Measurements are done at low frequency ($f=132.82$~Hz).
In the case of 3, 5, 7, and 9 terminals, the current vectors used in the experiment [see Eq.~\eqref{eq:analytic_current_vec}] are given by
\begin{align}
\mathbf{I_3} = \begin{pmatrix} 4816\\0\\ -942\end{pmatrix} \text{nA},
\quad 
\mathbf{I_5} = \begin{pmatrix} 5000\\0\\ -981\\ 0\\ 192
\end{pmatrix} \text{nA}, \nonumber \\
\\
\mathbf{I_7} = \begin{pmatrix} 4240\\0\\-832\\0\\162\\0\\ -32\end{pmatrix} \text{nA} \nonumber,
\quad
\mathbf{I_9} = \begin{pmatrix} 4050\\0\\-794\\0\\156\\0\\-31\\0\\6\end{pmatrix}\text{nA}.
\end{align}

\begin{figure*}[ht]
        \includegraphics[width=17.8cm]{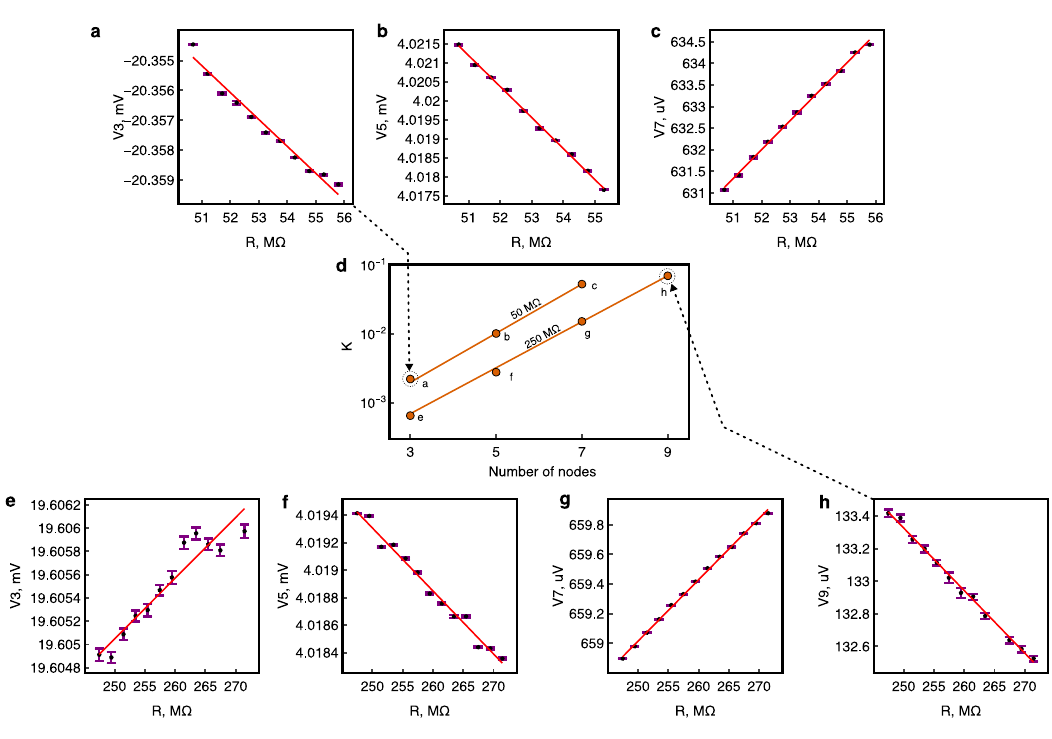}
\caption{\label{Fig. S1}
\textbf{Voltage dependencies used for the relative slope calculations.}
\textbf{(a-c)} Voltage on the last node versus the resistance $R \approx 50$~$\si{\mega\ohm}$ for 3, 5, 7-nodes systems. The red lines show linear fits that are used to calculate the relative slopes K in panel (d).
\textbf{(d)} Relative slope versus different number of nodes for $R \approx 50$~$\si{\mega\ohm}$ and for $R \approx 250$~$\si{\mega\ohm}$. 
The measurements are realized on a range of resistance corresponding to $10\%$ of $R$. 
Solid lines correspond to linear fits.
\textbf{(e-h)}~Voltage on the last node versus the resistance $R \approx 250$~$\si{\mega\ohm}$ for 3, 5, 7, and 9-nodes systems.}
\end{figure*}

Figure~\ref{Fig. S1} shows the voltage data used to calculate the values of K for various system sizes and ranges of $R$ that were shown in  \fig{Fig. 3}a.

\begin{figure*}[ht]
        \includegraphics[width=17.8cm]{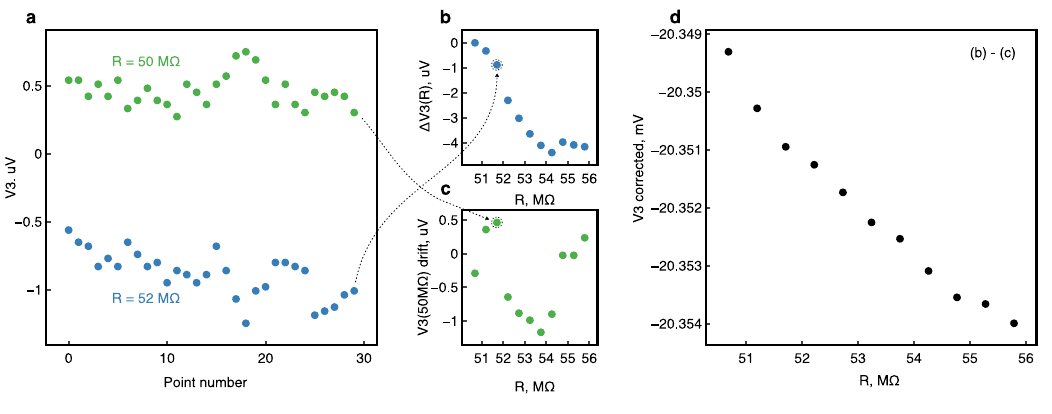}
\caption{\label{Fig. S2}
\textbf{Measurement steps for the last terminal of the 3-node device.}
\textbf{(a)} 30~experimental voltage points on the last terminal. Blue points correspond to $R = 52$~M$\Omega$. Green points were measured with a reference resistance ($R = 50$~M$\Omega$ here) just before the blue ones in order to compensate for the slow voltage drift.  
\textbf{(b)}~Voltage dependency on the last terminal versus resistance $R$. 
$\Delta V$ refers to an offset set at the first point.
\textbf{(c)} Slow voltage drift on the last terminal with $R = 50$~M$\Omega$.
\textbf{(d)}~Corrected voltage on the last terminal used to calculate the value of $K$ (same as \fig{Fig. S1}a). They are obtained by subtracting the data in panel (c) from the data in panel (b). 
Values are shown without the offset.  }
\end{figure*}

Figure~\ref{Fig. S2} demonstrates the measurement steps involved in obtaining \fig{Fig. S1}a. 
For each $R$, the voltage is calculated as an average over $30$~data points taken $1.5$~seconds apart for a $300$~ms integration time. This averaging process is shown in panel \fig{Fig. S2}a, which highlights the data points used to obtain one of the points in panel \fig{Fig. S2}b. Additionally, to take into account the slow voltage drift that occurs during the time span of the measurement, we measure the voltage with a reference resistance (50~M$\Omega$, 200~M$\Omega$ or open circuit for the different measurements shown in this work) just after reading the voltage with $R$. 
This drift voltage is shown in \fig{Fig. S2}c and is subtracted from the data shown in panel \fig{Fig. S2}b data to obtain the values displayed in \fig{Fig. S2}d, which is then used to calculate the value of $K$ for the given range of $R$.

The same measurement technique was used to measure the voltages on a wider range of $R$ in \fig{Fig. 2}c, d, and for the relative slope $K$ shown in \fig{Fig. 3}b. The only difference is the reference the resistance used to measure the slow drift ($R \to \infty$, no resistance attached between the first and last terminals).

\end{document}